\documentclass[aps,prx,ams,amsmath,singlecolumn,superscriptaddress,showkeys,preprint]{revtex4-2}

\usepackage{blindtext}
\usepackage{graphicx}
\usepackage{graphics}
\usepackage{epsfig}
\usepackage{amsmath}
\usepackage{amssymb}
\usepackage{amsfonts}
\usepackage{color}
\usepackage{bbm}
\usepackage{hyperref}
\usepackage[capitalise]{cleveref}
\usepackage{dcolumn}
\usepackage{bm}
\usepackage{caption}

\makeatletter
\def\@hangfrom@section#1#2#3{\@hangfrom{#1#2}#3}
\def\@hangfroms@section#1#2{#1#2}

\begin{document}

\title[Article Title]{Tunable reciprocal and nonreciprocal contributions to 1D Coulomb Drag}


\author{Mingyang Zheng}
    \affiliation{Department of Physics, University of Florida, Gainesville, FL 32611, USA}
\author{Rebika Makaju}
        \affiliation{Department of Physics, University of Florida, Gainesville, FL 32611, USA}
\author{Rasul Gazizulin}
        \affiliation{Department of Physics, University of Florida, Gainesville, FL 32611, USA}
        \affiliation{National High Magnetic Field Laboratory High B/T Facility, University of Florida, Gainesville, 32611, Fl, USA}
\author{Sadhvikas J. Addamane}
   \affiliation{Center for Integrated Nanotechnologies, Sandia National Laboratories, Albuquerque, NM 87185, USA}
\author{D. Laroche}
    \altaffiliation{\textbf{Email of Author to whom correspondence should be addressed:} dlaroc10@ufl.edu}
    \affiliation{Department of Physics, University of Florida, Gainesville, FL 32611, USA}
    





\maketitle

\section*{Abstract}

Coulomb drag is a powerful tool to study interactions in coupled low-dimensional systems. Historically, Coulomb drag has been attributed to a frictional force arising from momentum transfer whose direction is dictated by the current flow. In the absence of electron-electron correlations, treating the Coulomb drag circuit as a rectifier of noise fluctuations yields similar conclusions about the reciprocal nature of Coulomb drag. In contrast, recent findings in one-dimensional systems have identified a nonreciprocal contribution to Coulomb drag that is independent of the current flow direction. In this work, we present Coulomb drag measurements between vertically coupled GaAs/AlGaAs quantum wires separated vertically by a hard barrier only 15 nm wide, where both reciprocal and nonreciprocal contributions to the drag signal are observed simultaneously, and whose relative magnitudes are temperature and gate tunable. Our study opens up the possibility of studying the physical mechanisms behind the onset of both Coulomb drag contributions simultaneously in a single device, ultimately leading to a better understanding of Luttinger liquids in multi-channel wires and paving the way for the creation of energy harvesting devices.

\section*{Introduction}\label{sec1}

In recent years, there has been renewed interest in the study of one-dimensional systems (1D), owing to their potential for hosting topologically protected Majorana-Bound-States \cite{Lutchyn_2018, Flensberg_2021} and for realizing heat harvesting devices \cite{Sothmann_2014, Bhandari_2018}. Indeed, due to the strong electron-electron interactions arising from the significantly reduced screening \cite{voit_one-dimensional_1995, imambekov_universal_2009}, 1D systems are a rich platform for the observation of exotic quantum phases and states. The Tomonaga-Luttinger liquid (TLL) framework \cite{Tomonaga_1950, luttinger_exactly_1963, haldane_effective_1981} has been particularly successful in describing 1D systems and their properties such as universal power-law scaling in 1D arrays \cite{Tarucha_1995, bockrath_luttinger-liquid_1999, Sato_2019, wang_one-dimensional_2022} as well as spin-charge separation \cite{auslaender_spin-charge_2005, jompol_probing_2009} and charge fractionalization \cite{steinberg_charge_2008} in tunnel-coupled 1D wires.  

However, quantum transport in Coulomb-coupled TLLs is not yet clearly established. When two conductors are placed in close proximity, current sourced in the active conductor can induce a voltage in the passive conductor through Coulomb interactions, as illustrated in Fig. \ref{fig1}a. This phenomenon, called Coulomb drag, has historically been understood in terms of momentum transfer (MT) between the active and the passive circuit charge carriers \cite{Pogrebinskii_1977, gramila_mutual_1991, narozhny_coulomb_2016}, following the kinetic equations.  Within the kinetic theory approach, Coulomb drag is akin to a frictional effect where electrons in the drive wire scatter through the Coulomb interaction with electrons in the drag wire, imparting them with a fraction of their momentum. In ballistic systems, this effect follows the Onsager relations \cite{Onsager_1931}, implying a reciprocal drag voltage $V_{drag}$ upon both layer exchange and current direction reversal. For standard electron-electron scattering, the polarity of the drag voltage is negative, as the negatively charged electrons are dragged along the current flow. 

Another physical interpretation comes from the picture of current rectification (CR) \cite{Kamenev_1995, Borin_2019}, which interprets Coulomb drag as a rectification of energy, or noise fluctuations by the passive circuit. Both interpretations yield identical results for weak electronic interlayer correlations within the domain of linear response and crucially rely on the presence of an intrinsic electron-hole asymmetry, preventing the cancellation of electron-electron and electron-hole contributions \cite{Kamenev_1995}. The strong electronic correlations inherent to TLLs break down this equivalence, and both the MT approach \cite{klesse_coulomb_2000, pustilnik_coulomb_2003, fiete_coulomb_2006, dmitriev_coulomb_2012} and the CR approach \cite{nazarov_current_1998, levchenko_coulomb_2008} yield distinct predictions for 1D Coulomb drag. In particular, the MT models do not explicitly account for fluctuating electrical fields, nor for the translational symmetry breaking that can occur in disordered mesoscopic circuits. They therefore predict a reciprocal Coulomb drag signal.  Alternatively, CR models, which are inherently non-linear, explicitly account for symmetry breaking in mesoscopic systems. As such, they can exhibit a nonreciprocal drag signal, whose polarity is fixed by the microscopic details of the impurity potential in the mesoscopic wires.\\

CR models rely on the rectification in the passive layer of the electrical fields and their fluctuation originating from the active layer \cite{Kamenev_1995, narozhny_mesoscopic_2000, Levchenko_2008}. This process consists of two distinct contributions: one from the current-inducing electrical field, which transfers momentum with fixed energy and direction to the passive layer electrons, and another arising from a fluctuating electrical field.  This latter contribution imparts momentum to electrons in the passive layer at arbitrary energies and in arbitrary directions. Under particle-hole and translational invariance, this fluctuation-induced contribution averages out, resulting in a null Coulomb drag signal. However, in mesoscopic systems with finite disorder, a second order rectified drag current is generated in the passive layer when it is subjected to space and time dependent electrical fields, \emph{i.e.} when disorder and confinement potential non-uniformity break electron-hole and translational symmetry \cite{Kamenev_1995}. These nonreciprocal fluctuating contributions can be sizable and exceed the current-induced contribution \cite{narozhny_mesoscopic_2000}. Although the current-induced contribution may also lead to a nonreciprocal signal for broken translational invariance, it is the only contribution that can give rise to a reciprocal Coulomb drag signal. 

The reciprocal linear contribution is what is generally calculated within the MT framework, and  considers the contributions from large angle backscattering \cite{klesse_coulomb_2000, fuchs_coulomb_2005}, forward scattering \cite{pustilnik_coulomb_2003, Aristov_2007} and small-angle backscattering \cite{dmitriev_coulomb_2012, Dmitriev_2016}. Notably, these models all predict that, at sufficiently low temperature, large angle backscattering dominates, leading to the creation of a charge-density wave between both wires and an exponentially diverging drag resistance with decreasing temperature. MT models explicitly take into consideration the strength of electron-electron interactions in determining the temperature dependence of the drag signal, yielding a rich non-monotonic temperature dependence of the drag signal. In contrast, while the CR models take into account fluctuations and symmetry breaking, current work in the literature doesn't explicitly account for strong electron-electron interactions, resulting in a predicted quadratic temperature dependence of those models \cite{narozhny_mesoscopic_2000, Levchenko_2008}. In previous 1D drag experiments, the distinction between current-inducing and fluctuating electrical fields was not explicitly addressed, and some experiments were consistent with reciprocal drag signals \cite{debray_experimental_2001, yamamoto_negative_2006, laroche_1d-1d_2014, du_coulomb_2021} while others exhibited nonreciprocal drag signals \cite{yamamoto_hydrodynamic_2012, Tabatabaei_2020, anderson_coulomb_2021, makaju_nonreciprocal_2024}. Thus, a clear characterization of the parametric dependence and of the relative strength of reciprocal and nonreciprocal 1D Coulomb drag remains lacking. Such a study could provide crucial information on the energy-dependence of drag relevant parameters, such as the strength of electron-electron interaction $k_{c^{-}}$, paving the way for the realization of heat harvesting devices \cite{Sothmann_2014, roche_harvesting_2015} in one-dimensional geometry.\\

In this work, we report Coulomb drag measurements between 1D quantum wires in a vertically-coupled architecture \cite{laroche_positive_2011, laroche_1d-1d_2014} with simultaneous reciprocal and nonreciprocal contributions, allowing a clear comparison between these two drag-inducing paradigms. The relative strength of both contributions is tunable with both gate voltage and temperature. Both contributions exhibit non-linear signals, which are contrasted with each other and with theoretical predictions. We also conduct a systematic temperature-dependence analysis of 1D Coulomb drag and observe an upturn behavior at numerous gate positions. The functional shape of the temperature-dependent signals appears to follow either a power-law or an Arrhenius dependence, as typically predicted in 1D systems with strong electron interactions \cite{klesse_coulomb_2000, fuchs_coulomb_2005,dmitriev_coulomb_2012, chou_localization_2019, zhou_coulomb_2019, makaju_nonreciprocal_2024}. These results are contrasted with measurements in laterally-coupled quantum wires \cite{makaju_nonreciprocal_2024} to highlight the role of interwire separation in the onset of this control.\\

\section*{Results}

\subsection*{Device operation and wires characterization}\label{sec2}

A schematic of the vertically-coupled quantum wires devices used in this work is shown in Fig. \ref{fig1}b, with a typical optical image provided (see Supplementary Figure 1). As the two wires are defined in two separate quantum wells, they are only separated by a 15 nm-wide hard AlGaAs barrier grown by molecular beam epitaxy, resulting in an interwire separation $d_{vert} = 33$ nm. Each wire is defined by a pinch-off (PO) gate and a plunger (PL) gate, shown in gold. Fig. \ref{fig1}a shows the conducting region of the device when appropriate negative voltages are applied to all four gates, creating two independently contacted quantum wires. Crucially, this design allows for interlayer interactions only in the region where the two quasi-1D wires overlap. The PO gate design ensures that conduction in the top (bottom) layer can only occur on the right (left) side of the device, as highlighted by the continuous pink (blue) layer, representing the top (bottom) 2DEG conducting region. Two nominally identical devices fabricated on the same GaAs/AlGaAs heterostructure are presented in this study. For vertical device 1, the conductance of the top wire as a function of the top PL gate (TPL) for different bottom PL gate (BPL) voltages is displayed in Fig. \ref{fig1}c. The 1D subband plateau-like features are prominently visible in the smoothed data, while the raw data still reveal the presence of resonances, likely due to disorder-induced quantum dots. The bottom wire exhibits sharper subbands and fewer defects compared to the top wire, and is notably less resistive. Therefore, in subsequent Coulomb drag measurements, we utilize the top wire as the drive wire and the bottom wire as the drag wire, as the drag signal is more dependent on the drag wire quality. Coulomb drag measurements with the bottom wire as the drive wire are presented in Supplementary Note 3. The wires characterization of device 2 is presented in Supplementary Note 11. Despite the identical heterostructure and fabrication process, device 2 shows an enhanced conductance compared to device 1, implying a lower disorder level.\\

\subsection*{Coulomb drag measurements}\label{sec3}

First, we investigate the subband dependence of the drag signal at the base temperature of the cryostat, with an electron temperature below 15 mK. Fig. \ref{fig2}a and Fig. \ref{fig2}b present the Coulomb drag mapping as a function of both TPL and BPL gate voltages. In this measurement, the drag signal is recorded using two different drive current directions: Fig. \ref{fig2}a displays the drag signal $V_{drag}^R$ where the drive current direction aligns with the drag voltage direction, while Fig. \ref{fig2}b shows $V_{drag}^L$ where the drive current direction is reversed. We note that the drag voltage direction is merely a sign convention for the voltage measurement, and this convention remains fixed upon reversal of the drive current direction. Clear vertical stripes corresponding to the drag wire subband positions are observed in both directions. Additionally, the drag signal exhibits sign changes at various positions, consistent with prior experiments \cite{laroche_positive_2011, makaju_nonreciprocal_2024}. Alongside the vertical stripes, sloped stripes are observed, corresponding to constant density lines of the drive wire. To investigate the reciprocity of the drag signal, we extract the symmetric component, $V_{drag}^{S} = \frac{V_{drag}^R + V_{drag}^L}{2}$, and the antisymmetric component $V_{drag}^{AS} = \frac{V_{drag}^R - V_{drag}^L}{2}$. As the nonreciprocal component's polarity remains unchanged upon reversing the drive current direction, it is given by the symmetric contribution while the reciprocal component is given by the antisymmetric contribution. We note that $V_{drag}^{L,R}$ is the AC response to the small, low-frequency AC current sourced through the drive wire. Up to a sign convention, dividing this AC drag voltage by the sourced AC current (typically 2 nA unless stated otherwise) yields the drag resistance $R_{drag}$ typically reported in Coulomb drag experiments. To assess the relative strengths of these components, we plot in Fig. \ref{fig2}c the ratio between the symmetric and antisymmetric components, $r_{SAS} =  \frac{V_{drag}^{S}+\Gamma}{V_{drag}^{AS}+\Gamma} $. Here an offset $\Gamma  \sim 0.02 \mu$V, comparable to the measurement noise, is added to the ratio to distinguish nearly null signals (ratio near 1) from signals where the antisymmetric component is strongest (ratio near 0). At base temperature, the nonreciprocal component (ratio larger than 2) dominates most gate-space regions with a few notable exceptions where the primary drag contribution is reciprocal. However, the antisymmetric component dominates across the entire map at approximately 800 mK, as shown in Fig. \ref{fig2}d, providing clear evidence of the distinct temperature dependence of the two components. We also measured the two contributions to the drag signal in a lateral device, wherein the two wires are defined in a single quantum well and are separated by a 150 nm-wide gate-defined potential barrier, resulting in an interwire separation $d_{lat} \geq 250$ nm. Fig. \ref{fig2}e and Fig. \ref{fig2}f illustrate the ratio 2D map at base temperature (electron temperature around 100 mK) and 800 mK. In the lateral device, the reciprocal component exhibits fewer regions of dominance, with their range only expanding slightly at 800 mK. Thus, both vertically and laterally-coupled devices demonstrate tunable contributions of Coulomb drag through gate and temperature adjustments, but that tunability is notably stronger in vertically-coupled devices with a smaller interwire separation.\\

A hallmark of the energy fluctuation models is their predicted nonlinear current-voltage relations. While near-equilibrium thermal fluctuations in the drive circuit are predicted to generate a drag signal linearly proportional to the drive current, shot-noise rectification is predicted to exhibit a quadratic drive current dependence \cite{Levchenko_2008}. Higher order contributions can also arise when considering the impact of odd cumulants of the current noise \cite{Borin_2019}. To investigate the nonlinear regime of the drag signal, the current-voltage characteristics of the drag signal are presented in Fig. \ref{fig3}, both for the reciprocal and the nonreciprocal contributions. For all 4 gate configurations presented, the drag signal exhibits notable nonlinearity and non-monotonicity. The I-V relationship is reasonably well fitted by a 3rd order polynomial, as shown by the dotted lines in Fig. \ref{fig3}. Both the reciprocal and the nonreciprocal fitting parameters, reported in Supplementary Tables 1 and Table 2 of the Supplementary Information, are dominated by a linear term at low current, which is generally between 1 and 2 orders of magnitude stronger than the quadratic term and between 3 and 4 orders of magnitude stronger than the cubic term, in units of $\mu V$ and nA. These results, consistent with studies of the nonreciprocal drag in laterally-coupled quantum wires \cite{makaju_nonreciprocal_2024}, highlight the nonlinear nature of both the reciprocal and the nonreciprocal drag signal, as expected within the charge fluctuation model. However, the appearance of oscillations superimposed on a monotonic nonlinear power-law background has also been predicted for MT-induced reciprocal 1D Coulomb drag, owing to interactions with plasmon standing waves \cite{nazarov_current_1998, peguiron_temperature_2007}. In fact, such a model applied to the reciprocal component of the drag signal has been shown to yield self-consistent values for the interaction Luttinger liquid parameter $K_{c}$ \cite{zheng_quasi-1d_2024} from both the power-law dependence of the background and the oscillation period, given by $eV_{drive} = 2 \pi \hbar \omega_{L}$. Here, $\omega_{L} = \frac{v_{F}}{K_{c} L}$ is the plasmon frequency. However, for partially filled subbands, such an analysis is complicated by the difficulty in determining the 1D Fermi velocity $v_{F}.$ Additional theoretical work would also be required to confirm whether similar plasmon interactions can also play a role in the nonlinear regime of CR Coulomb drag, potentially explaining some of the discrepancies observed between the 3rd order polynomial fit and our data, as observed for the nonreciprocal drag component presented in Fig. \ref{fig3}a and b for instance. \\

The tunability of the two contributions from Coulomb drag in vertically-coupled wires enables us to study their temperature dependence individually and simultaneously on the same device. We present such a study in Fig. \ref{fig4} for device 1 at a constant drive wire density, selected to be at TPL = -0.735 V. Similar results extracted at different TPL values and at fixed BPL are presented in Supplementary Note 5, and yield qualitatively similar results. Fig. \ref{fig4}a and Fig. \ref{fig4}b illustrate the BPL gate-temperature 2D plots for the symmetric and antisymmetric drag components, respectively. Their ratio is presented in Fig. \ref{fig4}c. A similar ratio for device 2 is presented in Fig. \ref{fig4}d. Two distinct temperature regimes can be identified: a low-temperature regime below $\sim 1.5$ K where both components exhibit BPL gate tunable sign changes in the drag voltage polarity and a high-temperature regime where the drag signal is strictly positive, increases with increasing temperature, and is dominated by the antisymmetric component. In the low-temperature regime, the sign change of $V_{drag}^{S}$ overlaps the onset of drag wire subbands (see Supplementary Note 4), while $V_{drag}^{AS}$ is primarily positive, with only a few regions with negative $V_{drag}^{AS}$ and no clear correlation to the 1D wire subbands. The polarity change of $V_{drag}^{S}$ has recently been reported for nonreciprocal Coulomb drag \cite{makaju_nonreciprocal_2024}, and has also been reported for 1D Coulomb drag in the presence of a single quantum dot in one of the wires \cite{shimizu_negative_2005}. All these observations are consistent with charge fluctuation models for Coulomb drag in mesoscopic systems \cite{narozhny_mesoscopic_2000, Levchenko_2008}, which predict a Coulomb drag polarity that depends on the microscopic details of the wire and can easily induce negative Coulomb drag in the presence of translational asymmetry in the wire, favoring transmission in one direction over the other.  We note that while qualitatively similar results are observed in device 2, reciprocal Coulomb drag is present over a larger section of the system.  This is also consistent with the microscopic energy fluctuation model, as more highly transmitting wires are expected to have fewer defects, and hence a lesser extent of translational symmetry breaking. The fact that reciprocal Coulomb drag becomes predominant in the high temperature regime is also consistent with impurity-induced translational symmetry breaking.  Indeed, common impurities in GaAs/AlGaAs heterostructures are shallow DX-centers, whose energy lies close to the conduction band. As such, the notable relative increase in reciprocal drag signal at elevated temperatures suggests that the disorder-induced potential non-uniformity in our system is of the order of $\sim 1.5$ K. \\   

The transition between both regimes is characterized by an upturn in the magnitude of the drag signal in the vicinity of $T \sim 1.5$ K, both for symmetric and antisymmetric contributions. This transition is more easily visible in Fig. \ref{fig4}e-h, where $V_{drag}^{S}$ and $V_{drag}^{AS}$ are presented at different BPL values. Such an upturn had been previously predicted to occur due to the creation of a charge-density wave for positive drag resistance occurring through backscattering \cite{klesse_coulomb_2000, fuchs_coulomb_2005}, albeit at much lower temperatures, in the spin-incoherent regime \cite{fiete_coulomb_2006} or for negative drag arising from Umklapp scattering \cite{chou_localization_2019}. However, since this upturn occurs at the same temperature for both the reciprocal and the nonreciprocal contributions and for virtually all drag and drive wire densities, it is unlikely that the aforementioned models accurately describe this observation. In a typical quantum wire, 1D subbands typically become smeared out and non-visible through conductance measurements around $T \sim 1.5$ K, where the typical subband energy spacing becomes comparable to the temperature \cite{thomas_interaction_1998}. Combined with the disappearance of the subband structure in the reported Coulomb drag data above the upturn temperature, it is likely that the transition between the high and the low-temperature drag regimes arises from a fundamental change in the drag-inducing scattering mechanisms between separated and temperature-mixed 1D subbands.\\

\subsection*{Coulomb drag high-temperature regime}\label{sec4}

Fig. \ref{fig4}e-h, highlights the clear difference in the temperature dependence of reciprocal and nonreciprocal drag in the high-temperature regime. To elucidate this discrepancy, we analyzed the functional shape of the Coulomb drag temperature dependence.  Both contributions are well-described by a power-law ($V_{drag} \propto A \times T^{B}$), in qualitative agreement with numerous theoretical models for 1D Coulomb drag at elevated temperatures \cite{klesse_coulomb_2000, fiete_coulomb_2006, zhou_coulomb_2019}. The parameters extracted from a power-law fit in device 1 over the linear range of a log-log plot are presented in Fig. \ref{fig5}b, c, d, and e for four line cuts at TPL = -0.5475 V, TPL = -0.6225 V, TPL = -0.6788 V, and TPL = -0.735 V, respectively. Typical examples of the temperature dependence of the drag signal in a log-log plot are shown in Fig. \ref{fig5}a.  The power-law exponent of each drag component oscillates with changes in BPL gate voltage. This result is expected as, in 1D, the value of this exponent is predicted to depend on the strength of electron-electron interactions, which depends itself on the wires' electronic density \cite{klesse_coulomb_2000}. Similar oscillations are also observed in the drag magnitude. These gate-dependent peaks likely arise from a combination of factors, including the drag signal enhancement upon the opening of 1D subbands, the density dependence of the drag signal and the changes in the wires confinement potential and electrostatic screening  with increasing wire width.

For most gate positions, the power-law exponent of the reciprocal drag component, ranging from 3 to 5, exceeds that of the nonreciprocal component, ranging from 2.5 to 4. In addition, the power-law exponents from the nonreciprocal contribution exhibit minimal wire subband dependence. Such a discrepancy between the reciprocal and nonreciprocal regimes points towards either an energy dependence of the Luttinger liquid interaction parameters, or towards changes in scattering rate with the magnitude of the momentum transferred between the drive and the drag wire.  

Quantitatively, the nonreciprocal drag signal exhibits power-law exponents notably different than the quadratic temperature dependence predicted by non-interacting models \cite{levchenko_coulomb_2008}.  Further theoretical work, including strong electron-electron interactions within the Luttinger liquid framework, will likely be required to provide a better comparison to our experimental data. Within the MT formalism, numerous models have predicted a power-law dependence for reciprocal 1D Coulomb drag, albeit in the single-subband limit. For non-identical wires, density-independent power-law exponents of relatively large value, $B = 5$ \cite{pustilnik_coulomb_2003, Aristov_2007} or $B = 4$ \cite{pereira_spin_2010} have been predicted for forward-scattering induced drag. However, the reported power-law exponents clearly show density-dependent oscillations, even within a single subband, and as such do not readily match these predictions.  

For wires with minimal disorder and matched density, backscattering models predict a power-law exponent of $B =2K_{c} - 1$ \cite{klesse_coulomb_2000}. However, within this model, the extracted power-law exponent values lead to $K_{c} > 1$. Repulsive interactions for moderately separated quantum wires are expected to yield $K_{c} < 1$, a result that is incompatible with our observations. However, repulsive interactions can still yield $K_{c} > 1$ for wires in very close proximity such that the small momentum scattering coupling, within the g-ology framework \cite{voit_one-dimensional_1995}, are nearly equal \cite{klesse_coulomb_2000}. Additional theoretical work would be required to explore this possibility and verify whether matched density models apply to non-identical wires in the presence of disorder in the multi-subband regime. Models for Coulomb drag between two spin-incoherent TLLs \cite{fiete_coulomb_2006} predict a power-law exponent of $8K_c-3$ in the high-temperature regime when the Fermi energy is higher than the spin exchange energy. Although this model would yield a Luttinger interaction parameter of $K_c = 0.875$ for $B = 4$, consistent with repulsive electron-electron interactions for moderately separated wires, it is only applicable at low electron density where $n a_{B} <1$. Here, $n$ is the 1D electron density and $a_{B} = 10.2 nm$ is the GaAs Bohr radius. From our estimates, such a low density can only be achieved in the single subband regime, and is not consistent with our results obtained at larger density. Therefore, none of the theoretical reciprocal Coulomb drag models accurately reflect the large, density-dependent power-law exponents observed in the multi-subband regime.  In addition, neither of these models readily explains the discrepancy observed between the power-law exponents extracted from reciprocal and nonreciprocal Coulomb drag. This highlights the need for more theoretical work to grasp the richness of Coulomb drag between multi-channel quantum wires away from the ballistic regime.\\

Looking at the power-law magnitudes, both the symmetric and the antisymmetric drag components increase as the electron density decreases, consistent with enhanced interaction strength and reduced screening at lower electron density. Notably, this dependence is well-fitted by a linear slope going over small density oscillations, as shown by the solid lines in Fig. \ref{fig5}b-e. For vertical device 1, the slopes for the symmetric component are -14.3, -11.7, -17.2, and -19.9, respectively, while the ones for the antisymmetric components are -24.6, -28.6, -37.5, and -33.3. The ratio between these slopes, $r_{int}$, is presented in the inset of Fig. \ref{fig5}e. In the high-temperature regime where the 1D subband density of states is notably smeared, it is reasonable to assume that the quasi-1D density, and hence $\textbf{k}_{\textbf{F}}$, varies linearly with plunger gate voltage, similarly to what happens in 2D systems \cite{chen_2021}. Thus, the linear dependence of $\text{ln}(A)$ with gate voltage (or on $\textbf{k}_{\textbf{F}}$) implies that $A \propto \text{exp}(C\cdot \textbf{k}_{\textbf{F}})$, where the constant $C$ corresponds to the linear slope. This underlying exponential dependence of the drag signal upon $\textbf{k}_{\textbf{F}}$ had been predicted for MT-induced 1D Coulomb drag, where the drag magnitude $\lambda_{AS}\sim e^{-2\textbf{k}_{\textbf{F}}d}$ \cite{narozhny_coulomb_2016}. Since the ratios $r_{int}$ are in the vicinity of 2, it implies that the nonreciprocal Coulomb drag magnitude $\lambda_S\sim e^{-\textbf{k}_{\textbf{F}}d}$. The stronger $\textbf{k}_{\textbf{F}}d$ dependence of reciprocal Coulomb drag thus explains why CR Coulomb drag remains dominant at high temperatures in laterally coupled devices: the nearly 6 times larger interwire separation renders the reciprocal contribution negligible, even at elevated temperatures. This is also consistent with CR Coulomb drag models. Indeed, as fluctuation-induced Coulomb drag spans a wider energy range than current-induced drag, low-momentum contributions would decay more slowly than the contributions arising from momentum transfer at the Fermi energy, hence enabling the nonreciprocal drag signal to persist at larger interwire separations. We note that other theoretical models \cite{fuchs_coulomb_2005, dmitriev_coulomb_2012, Dmitriev_2016} predict an Arrhenius temperature dependence of the Coulomb drag signal ($V_{drag} \propto \alpha e^{(\frac{-\beta}{T})}$). In contrast to the aforementioned theories, the result of such an analysis, presented in Supplementary Note 7, shows negligible $k_{F}$ dependence and is not further discussed. We also note that comparable results have been observed in device 2, as highlighted in Fig. \ref{fig5}f, while the slightly noisier power law fitting parameters can be accounted for by the higher upturn temperature, resulting in fewer points in the high-temperature fitting range for device 2. Indeed, the values of the power-law exponent, both for reciprocal and nonreciprocal drag, as well as the discrepancy in the exponential decay of the magnitude of the reciprocal and nonreciprocal drag contributions, are very similar in both devices. Finally, while a similar analysis was performed in the low-temperature regime, we were unable to identify the functional temperature dependence of the signal, nor were we able to identify notable trends in the evolution of the fitting parameters when either a power-law or an Arrhenius dependence was considered. The reduced temperature range where the low-temperature signal is monotonic, combined with the increased impact of disorder on transport at low temperatures, explains the lack of a clear trend in the analysis. Additional details are provided in the Supplementary Note 9.   

\section*{Discussion}\label{sec6}

Overall, our study successfully demonstrated the gate and temperature tunability of reciprocal and nonreciprocal Coulomb drag contributions while contrasting their density and temperature dependences.  Interestingly, both components exhibited positive and negative polarities at temperatures below 500 mK, with the nonreciprocal component changing sign whenever the subband number changed. Notably, the observation of a negative reciprocal component of the drag signal sits at odds with Coulomb drag prediction in clean TLLs \cite{klesse_coulomb_2000, pustilnik_coulomb_2003, dmitriev_coulomb_2012} and highlights the role of disorder in our system, and its impact on the nature of Coulomb scattering. While a number of observations, such as a temperature upturn in the drag signal strength and an increasing drag magnitude with decreasing temperature, are consistent with Luttinger liquid physics, the specific values of the power-law exponents extracted are not readily explained within the current Coulomb drag literature. Our observations highlight the richness of interactions between Coulomb-coupled 1D systems and the need for further theoretical investigations, especially in the presence of disorder, detailing how momentum transfer away from $k_{F}$ affects the nonreciprocal Coulomb drag component compared to the reciprocal one. Such studies could be instrumental to the realization of novel applications in the realm of topological quantum computing \cite{Lutchyn_2018} or heat harvesting \cite{Sothmann_2014}.

\section*{Methods}\label{sec7} 

\noindent \textbf{Material growth and device fabrication}\\
The wires were patterned on an n-doped GaAs/AlGaAs electron bilayer heterostructure with two 18-nm-wide quantum wells separated by a 15-nm-wide Al$_{0.3}$Ga$_{0.7}$As barrier. The unpatterned density and mobility of the GaAs quantum well are $n = 2.98 \times 10^{11} \, \text{cm}^{-2}$ and $\mu = 7.4 \times 10^4 \, \text{cm}^2 / \text{V} \cdot \text{s}$, respectively. After a mesa-structure was wet-etched using phosphoric acid, Ge–Au–Ni–Au ohmic contacts were deposited on the structure and annealed at 420 $^{\circ}$ for 60 s. A set of two Ti–Au split gates was then defined on the surface of the heterostructure, using electron-beam lithography [Fig. \ref{fig1}a]. Once the upper side processing was complete, bare GaAs was epoxied on top of the heterostructure and the sample flipped, mechanically lapped and chemically etched using subsequent citric and hydrofluoric mixtures until the lower 2DEG was only $\sim$ 150 nm away from the lower surface (now on top of the device), following the EBASE technique \cite{weckwerth_epoxy_1996}. To ensure that no off-mesa leakage occurred between the bottom and top gates, a thin 40 nm layer of Al$_2$O$_3$ was deposited on the new surface using atomic layer deposition. Using phosphoric acid, vias were then etched through the surface to enable electrical connection to the ohmic contacts and the split gates buried under the surface of the device. Finally, using electron-beam lithography, another set of Ti–Au split gates was defined on the top side of the device, and aligned with the bottom gates. The end result is depicted in Fig. \ref{fig1}a, with an optical image provided in Supplementary Note 1.\\

\noindent\textbf{Measurement techniques}\\
Transport measurements were performed in a dilution refrigerator (LD250, Bluefors) with a base temperature less than 7 mK. The device was mounted in an experimental cell, which is thermally anchored to the mixing chamber of the dilution refrigerator. The polycarbonate cell is then filled with liquid Helium-3. The liquid is thermalized to the mixing chamber of a dilution refrigerator via annealed silver rods that enter the Helium-3 cell. These bring the system’s base temperature near that of the dilution unit, below 15 mK. All measurements were performed in an ultraquiet environment, shielded from electromagnetic noise. RC filters with cutoff frequencies of 50 kHz were employed to reduce RF heating. All measurements were performed using standard low-frequency lock-in ampliﬁcation techniques. Additionally, source-measure units were used to source and measure DC signals applied to the electrostatic gates. Transport measurements on individual quantum wires were performed at base temperature using a constant 100 $\mu$V excitation at 13 Hz in both wires in a two-contact configuration. The Coulomb drag measurements were performed in a constant-current mode where a typical 2 nA current was sent at 13 Hz through the drive wire. Some measurements were performed with a larger current, up to 9 nA. In this configuration, the out-of-phase current was always much smaller than the in-phase current. The tunneling measurements, as shown in Supplementary Note 2, were performed by sending a small source-drain voltage across the device for different top PO gate and bottom PO gate values. The combination was selected such that the tunneling resistance between the two wires was larger than 10 M$\Omega$ in a bias range of $\pm 1.5 \text{mV}$. Detailed configuration of the drag measurement is also presented in the Supplementary Note 2.



\section*{Data availability}\label{sec8} 

The data that support the findings of this study are available within the paper and its Supplementary Information. The raw data that support the findings of this study are available in Zenodo with the identifier: doi.org/10.5281/zenodo.15756744

\section*{Code availability}\label{sec9} 

The code used for analysis is available in Zenodo with the identifier: doi.org/10.5281/zenodo.15756744



\section*{Acknowledgments}\label{sec9} 


This work was performed, in part, at the Center for Integrated Nanotechnologies, an Office of Science User Facility operated for the U.S. Department of Energy (DOE) Office of Science. Sandia National Laboratories is a multimission laboratory managed and operated by National Technology $\And$ Engineering Solutions of Sandia, LLC, a wholly owned subsidiary of Honeywell International, Inc., for the U.S. DOE’s National Nuclear Security Administration under contract DE-NA-0003525. The views expressed in the article do not necessarily represent the views of the U.S. DOE or the United States Government. Part of this work was conducted at the Research Service Centers of the Herbert Wertheim College of Engineering at the University of Florida.  A portion of this work was also performed at the National High Magnetic Field Laboratory and was partially supported by the National High Magnetic Field Laboratory through the NHMFL User Collaboration Grants Program (UCGP). The National High Magnetic Field Laboratory is supported by the National Science Foundation through NSF/DMR-1644779 and NSF/DMR-2128556 and the State of Florida. Finally, we also acknowledge Alex Levchenko for enlightening discussions.

\section*{Author contribution}\label{sec10} 
M.Z., R.M., and D.L. fabricated the sample. M.Z. and R.G. performed the measurements of the vertically coupled devices. R.M. performed the measurements of the laterally coupled devices. S.J.A. performed the growth of the double quantum well heterostructures. D.L. designed and supervised the experiments. M.Z. analyzed the data. M.Z. and D.L. co-wrote the Letter, and all authors discussed the results and commented on the manuscript.

\section*{Competing interests}\label{sec11}

The authors declare no competing interests.


\begingroup
\begin{figure}[h!]
\centering
\includegraphics[width=0.8\textwidth]{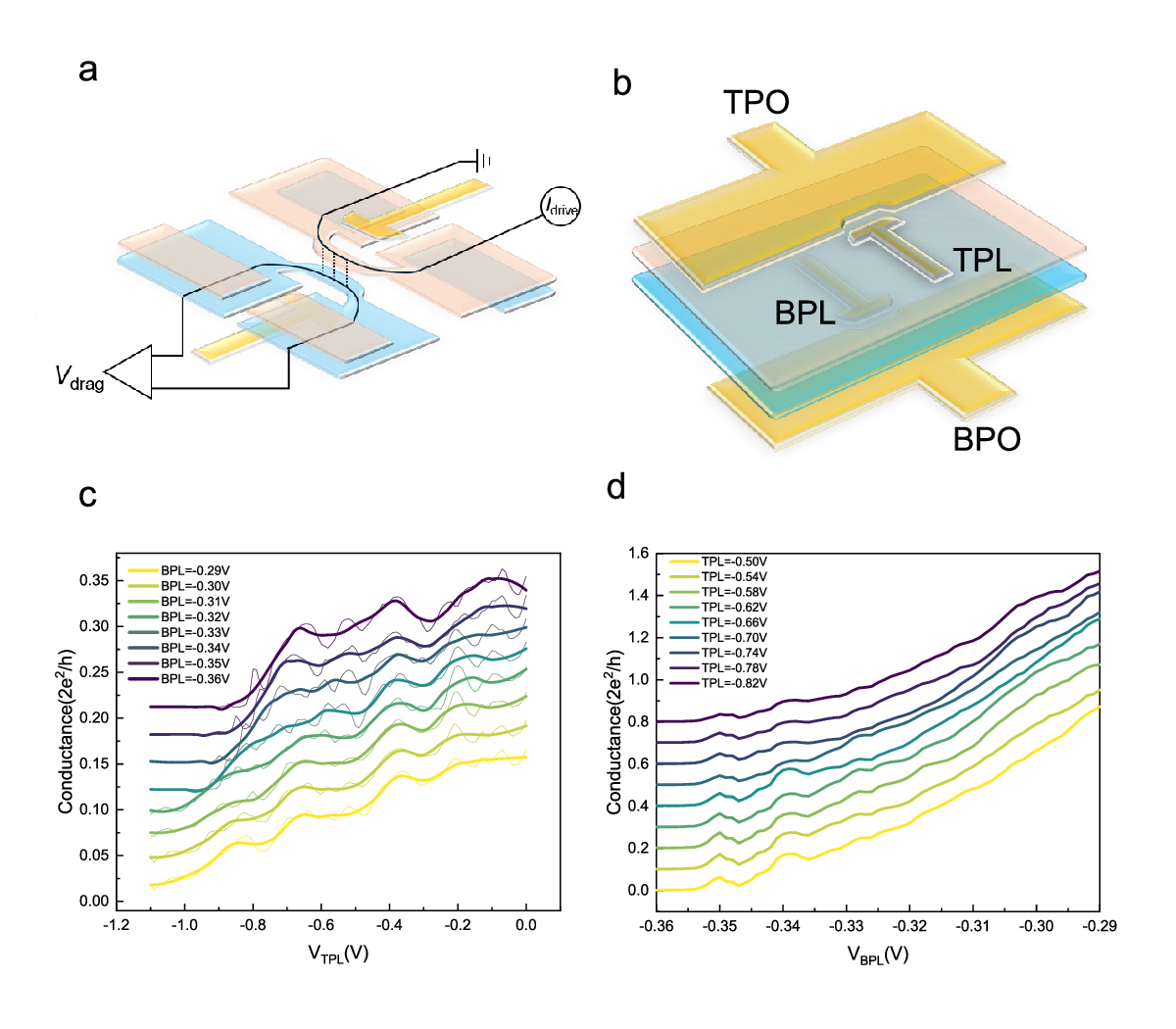}

\caption{\scriptsize{\textbf{Design of the vertically integrated quantum wire device.} \textbf{a}, Vertically aligned quantum wires are defined electrostatically in a double-layer two-dimensional electron gas (2DEG) system. Negative voltages applied to four surface gates shape the conducting regions, with the top-layer 2DEG shown in pink and the bottom-layer 2DEG in blue. The white section represents sections where the electron density has been reduced to the point that the 2DEG is insulating. As such, conduction across the top (bottom) layer can only occur on the device's right (left) side. Plunger gates, shown in gold, enable control of the wire's width, or subband occupancy. Pinch-off (PO) gates are omitted for clarity. \textbf{b}, Schematic of the active part of the double quantum wire device without any voltages applied. Each wire is defined by one plunger and one pinch-off gate on each 2DEG. The gates are shown in gold. The PO gates are primarily used to independently contact the quantum wires and minimize tunneling current between them, while the PL gates are used to adjust the wire's width and electronic density, similarly to previous work in vertically-coupled quantum wires \cite{laroche_positive_2011, laroche_1d-1d_2014}. In the interacting region of the device, two vertically-superimposed independent quantum wires are created, leveraging selective layer depletion with the PO gates. \textbf{c}, Top (drive) wire conductance as a function of the top (drive) and bottom (drag) gate voltages for vertical device 1. Plateau-like features are clearly visible in the smoothed data (thick line). Successive line-cuts are vertically offset by 0.03 (2e$^{2}$/h) for visibility, with the lowest BPL value at the bottom. \textbf{d}, Bottom (drag) wire conductance as a function of the bottom (drag) and top (drive) gate voltages for vertical device 1. Successive line-cuts are vertically offset by 0.1 (2e$^{2}$/h) for visibility purposes, with the lowest BPL value at the bottom. In both wires, the conductance plateaus are not quantized at integer values of 2e$^2$/h as the wires are non-ballistic. (Reprinted figure 1\textbf{a}, \textbf{b} with permission from M. Zheng, R. Makaju, R. Gazizulin, A. Levchenko, S. J. Addamane, and D. Laroche, Phys. Rev. Lett. 134, 236301 (2025). Copyright (2025) by the American Physical Society.)}}\label{fig1}

\end{figure}
\endgroup

\begingroup
\begin{figure}[h!]
\centering
\includegraphics[width=1\textwidth]{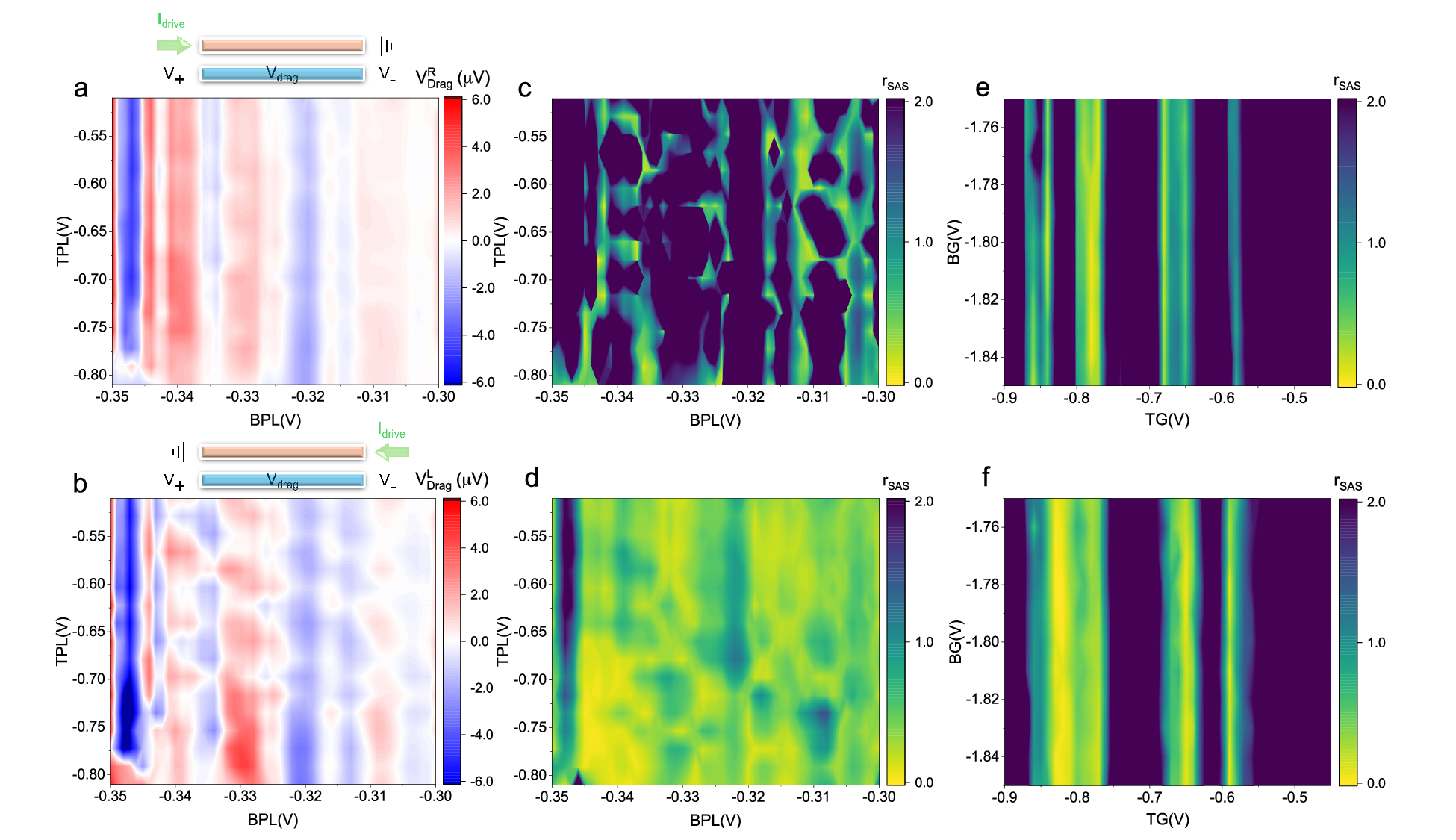}
\caption{\scriptsize{\textbf{Nonreciprocal drag in quantum wires.}  \textbf{a}, Drag voltage as a function of the top (drive) and bottom (drag) gate voltages for vertical device 1 when the drive current is in the same direction as the drag voltage measurement. The measurement was performed at the cryostat base electron temperature, below 15 mK. \textbf{b}, same as \textbf{a}, but with the drive current direction reversed. \textbf{c,d,} Ratio between symmetric and antisymmetric components (see text) of the vertically-coupled device at (\textbf{c}) the cryostat base electron temperature and (\textbf{d}) at 800 mK. \textbf{e,f} $r_{SAS}$ as a function of the bottom (drive) gate and top (drag) gate voltages for laterally-coupled quantum wires (horizontal device) at (\textbf{e}) 100 mK and  (\textbf{f}) 800 mK.}}\label{fig2}
\end{figure}
\endgroup

\begin{figure}[h]
\centering
\includegraphics[width=0.8\textwidth]{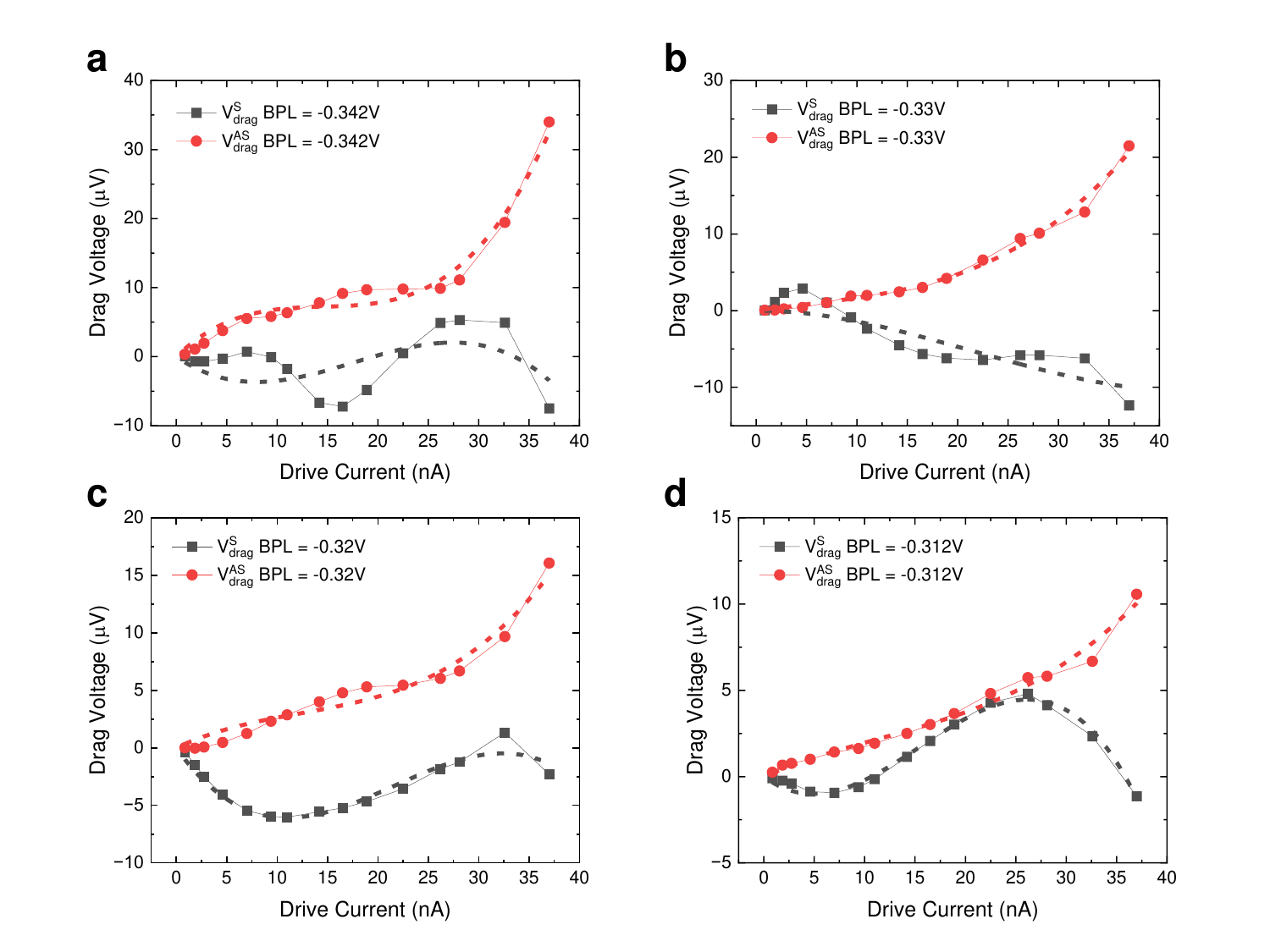}
\caption{\scriptsize{\textbf{Current-voltage characteristics of reciprocal and nonreciprocal drag.} Current-Voltage dependence of the reciprocal (red circles) and nonreciprocal (black squares) Coulomb drag signal at TPL = -0.69 V for \textbf{a} BPL = -0.342 V, \textbf{b} BPL = -0.33 V, \textbf{c} BPL = -0.32 V and \textbf{d} BPL = -0.312 V in vertical device 1. The data is fitted to a 3rd order polynomial, shown in the dotted line, whose parameters are reported in Supplementary Tables 1 and Table 2 of the Supplementary Information.
}}\label{fig3}
\end{figure}

\begin{figure}[h!]
\centering
\includegraphics[width=1\textwidth]{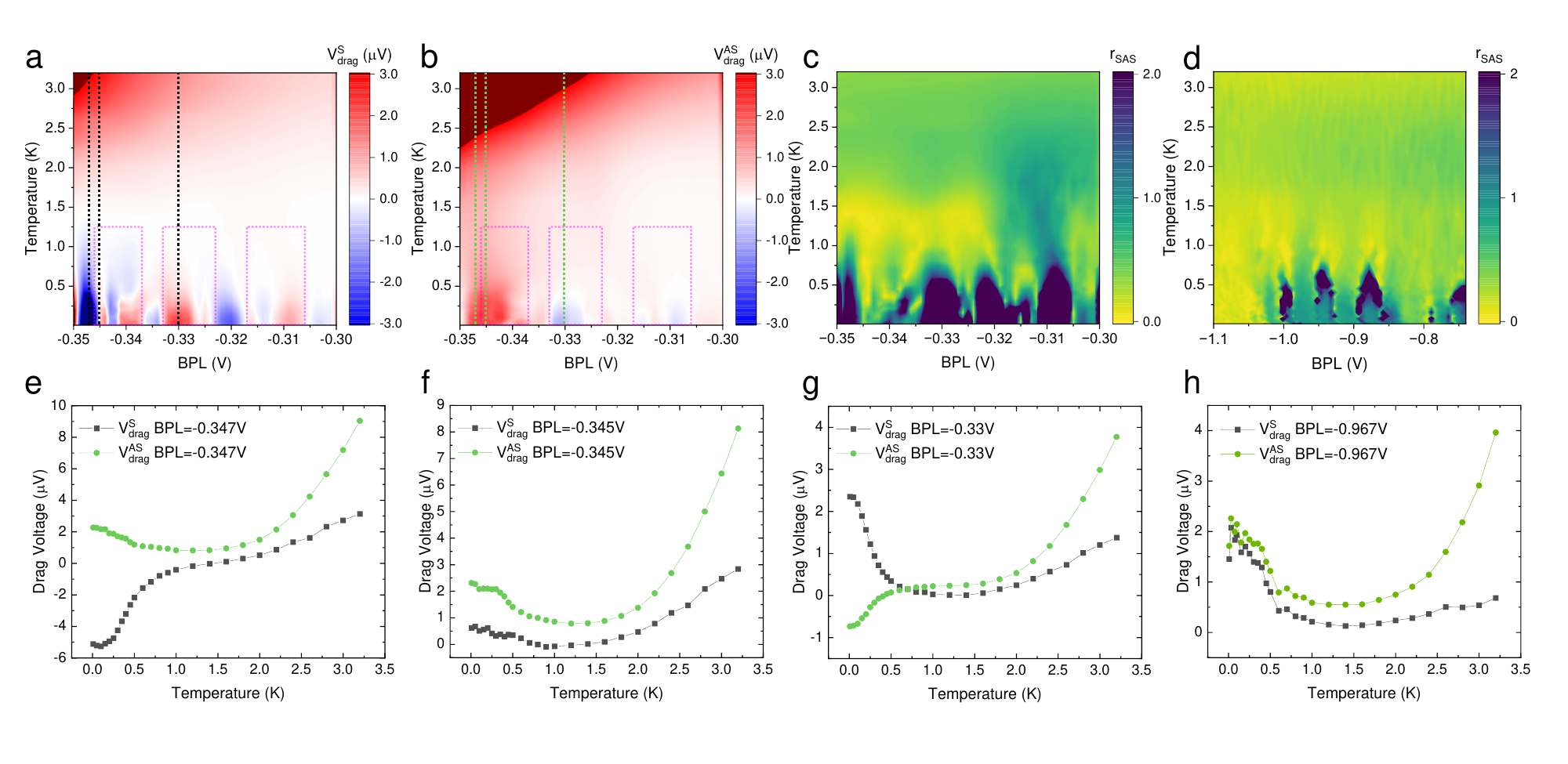}
\caption{\scriptsize{\textbf{Gate and temperature tunable rectified drag.} \textbf{a,b,} BPL gate-temperature 2D plot at TPL = -0.735 V of \textbf{a} the symmetric component and \textbf{b} the antisymmetric component of the drag signal in vertical device 1. The magenta dotted boxes represent the range of the first, second, and third subbands of the drag wire from left to right. \textbf{c}, The ratio between the symmetric and the antisymmetric drag components, $r_{SAS}$, as a function of BPL gate voltage and temperature. Above $\sim$ 750 mK, the drag signal is dominated by the antisymmetric component. \textbf{d,} $r_{SAS}$ at TPL = -0.410 V for vertical device 2, showing consistent dominance of the antisymmetric drag component above $\sim$ 750 mK. \textbf{e,f,g} The symmetric and antisymmetric drag components as a function of temperature taken in device 1 at constant TPL = -0.735 V for (\textbf{e}) BPL = -0.347 V, (\textbf{f}) BPL = -0.345 V and (\textbf{g}) BPL = -0.33 V, corresponding to the dotted lines in \textbf{a} and \textbf{b}. \textbf{h,} The symmetric and antisymmetric drag components as a function of temperature at BPL = -0.967 V and TPL = -0.410 V for vertical device 2.
}}\label{fig4}
\end{figure}

\begin{figure}[h!]
\centering
\includegraphics[width=1\textwidth]{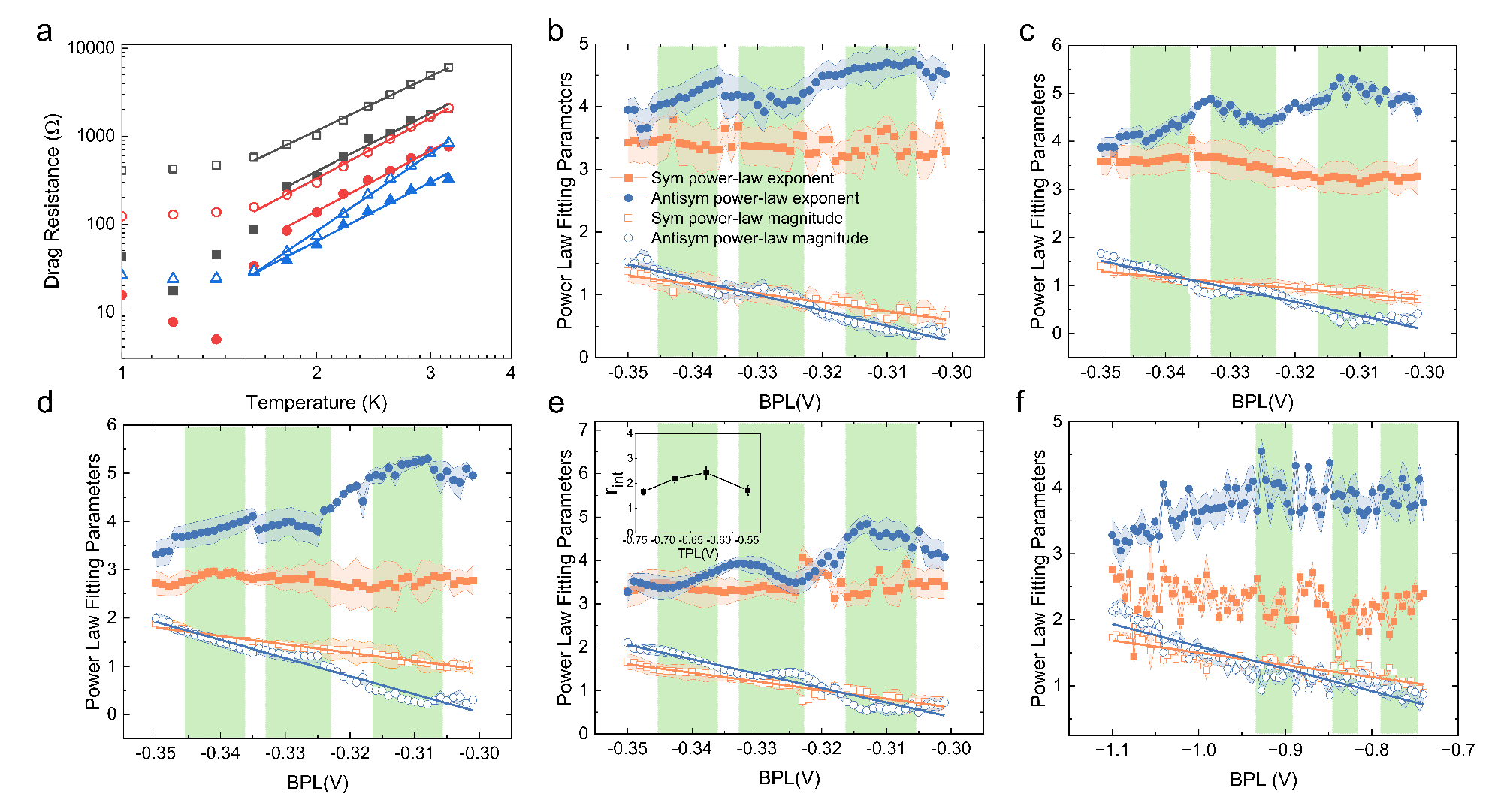}
\caption{\scriptsize{\textbf{High-temperature line cut analysis.} \textbf{a,} Log-log plot and power-law fitting for TPL = -0.735 V for the symmetric component $R_{drag}^{S}$ (full symbols) and the antisymmetric component $R_{drag}^{AS}$ (open symbols) of the drag resistance for vertical device 1. From top to bottom, the BPL gate voltages are -0.35 V (black), -0.33 V (red), and -0.31 V (blue). The solid lines are the linear fitted lines in the high-temperature range. The linearity of $R_{drag}$ in a log-log plot for $1.6$ K $\lesssim T \lesssim 3.2$ K confirms the power-law nature of the temperature dependence. The intercept of this fit gives the logarithm of the drag resistance magnitude while its slope gives the power-law exponent, since $\ln{(R_{drag})} = \ln{(A)} + B \times \ln{(T)}$ \textbf{b-e,} Power-law fitting results for the power-law exponent (full symbols) and magnitude (open symbols) at line cuts of (\textbf{b}) TPL = -0.5475 V, (\textbf{c}) TPL = -0.6225 V, (\textbf{d}) TPL = -0.6788 V and (\textbf{e}) TPL = -0.735 V for the symmetric (orange squares) and the antisymmetric (blue circles) drag components for vertical device 1. Similar to Fig. \ref{fig4}, the green shaded boxes represent the first, second, and third subbands of the drag wire from left to right, respectively. The error bars, arising from the endpoint selection in the fitting procedure, are shown as light shading. See Supplementary Note 6 for more details. The drag magnitude as a function of the BPL gate voltage has been linearly fitted, and the solid lines are the linear fits. Inset of (\textbf{e}): Ratio $r_{int}$ between the slopes of the symmetric and antisymmetric magnitudes. The errors in this data were calculated using a bootstrap Monte Carlo method. See Supplementary Note 10 for more details. \textbf{f,} Power law fitting results of the line cut TPL = -0.410 V of vertical drag device 2. Similar slopes and intercepts are obtained in both devices. In device 2, $r_{int} = 1.83$, also in agreement with the results from device 1. 
}}\label{fig5}
\end{figure}







\end{document}